\documentclass{aa}

\usepackage{txfonts}
\usepackage{graphicx}
\usepackage{epsf}

\begin{document}

\title{Far infrared observations of pre-protostellar sources in
       Lynds~183 
         \thanks{Based on observations with ISO, an ESA project
         with instruments funded by ESA Member States (especially the PI
         countries: France, Germany, the Netherlands and the United Kingdom)
         and with the participation of ISAS and NASA} }

\author{  K.\ Lehtinen\inst{1} \and K.\ Mattila\inst{1} \and 
          D.\ Lemke\inst{2} \and M.\ Juvela\inst{1} \and T.\ Prusti\inst{3}
          \and R.\ Laureijs\inst{4} }

\institute{Observatory, T\"ahtitorninm\"aki, P.O.\ Box 14, 00014
           University of Helsinki, Finland
           \and
           Max-Planck-Institut f\"ur Astronomie, K\"onigstuhl 17, D-69117 
           Heidelberg, Germany
           \and
           ESTEC/SCI-SAF, Postbus 299, NL-2200 AG Noordwijk, The Netherlands
           \and
           ISO Data Centre, ESA Satellite Tracking Station, Villafranca	
           del Castillo, P.O.\ Box 50727, E-28080 Madrid, Spain}

\date{Received / accepted}

\offprints{K.\ Lehtinen (kimmo.lehtinen@helsinki.fi)}

\titlerunning{ISOPHOT far infrared observations of L~183}

\authorrunning{K.\ Lehtinen et~al.}

\abstract{
Using ISOPHOT maps at 100 and 200\,$\mu$m and raster scans at 100,
120, 150 and 200\,$\mu$m we have detected four unresolved far-infrared
sources in the high latitude molecular cloud L~183. Two of the sources
are identified with 1.3\,mm continuum sources found by Ward-Thompson
et~al.\ (\cite{wthompson94}, \cite{wthompson00}) and are located near
the temperature minimum and the coincident column density maximum of
dust distribution.  For these two sources, the ISO observations have
enabled us to derive temperatures ($\sim 8.3$\,K) and masses
($\sim$1.4 and 2.4\,M$_{\sun}$). They are found to have masses greater
than or comparable to their virial masses and are thus expected to
undergo gravitational collapse. We classify them as pre-protostellar
sources.  The two new sources are good candidates for pre-protostellar
sources or protostars within L~183.
 }

\maketitle

\section{Introduction}

Low-mass stars are known to form within dark clouds. To study the
initial conditions for their formation, it is important to probe the
physical conditions of both molecular gas and cold dust in the deep
interiors of such clouds.  Dense cores have been identified inside
dark clouds, some probably being in the stage of contraction to form a
star. A pre-protostellar core is defined as the stage in which a
gravitationally bound core has formed in a molecular cloud, but no
central hydrostatic protostar exists yet within the core (see e.g.\
Ward-Thompson et~al.\ \cite{wthompson94}).  Pre-protostellar cores are
thought to be very cold.  Thus, in many cases they escaped detection
by IRAS, limited to $\lambda \le 100\,\mu$m.

The long wavelength and multi-filter capabilities of ISOPHOT aboard
the Infrared Space Observatory (ISO) (Kessler et~al.\
\cite{kessler96}), combined with its improved sensitivity and spatial
resolution over IRAS, have been utilized in studying the far-IR
emission of molecular clouds and pre-stellar and young embedded
stellar objects within them.  Analysis of their physical parameters is
facilitated by using the spectral energy distributions (SEDs) obtained
from ISOPHOT multi-filter photometry.

\subsection{Lynds 183}

The dark cloud/large globule L~183, frequently cited also as L~134
North, is a prototypical dense cold molecular cloud. Its visual
extinction is estimated to be $\sim 17^{\mathrm m}$ based on its
200\,$\mu$m optical depth (Juvela et~al.\ \cite{juvela02}).  So far
there is no evidence for associated newly born stars such as T~Tauri
stars or IRAS point sources. However, Martin \& Kun (\cite{martinkun})
have found a bona fide T~Tauri star and H$\alpha$ emission line star
near L~183. These two stars are located outside our maps.  Given its
short distance of $\sim$~110 pc (Franco\ \cite{franco}) L~183 provides
a good spatial resolution with the ISO FIR beam size (ISOPHOT's
spatial resolution of 45$\arcsec$ at 100\,$\mu$m corresponds to
0.02\,pc at the distance of L~183). The location at high Galactic
latitude ($b = 36$ deg) minimizes contamination with unrelated cirrus
along the line of sight. The location off the Galactic plane at $z =
65$\,pc implies that the impinging ultraviolet radiation field is
strongly asymmetrical.

A Digitized Sky Survey\footnote{The Digitized Sky Survey was produced
at the Space Telescope Science Institute under U.S.\ Government grant
NAG W-2166. The images of these surveys are based on photographic data
obtained using the Oschin Schmidt Telescope on Palomar Mountain and
the UK Schmidt Telescope. The plates were processed into the present
compressed digital form with the permission of these institutions} red
plate ('Equatorial Red' survey (UK Schmidt) IIIaF + RG610) image of
L~183 is shown in Fig.~\ref{optIR}. In Fig.~\ref{optIR}b, the original
image has been smoothed with a FWHM=9\arcsec (nine image pixels)
Gaussian kernel.  The cloud has a dark core - bright rim structure,
which is characteristic of clouds having high optical depth at visual
wavelengths that are illuminated by the diffuse interstellar radiation
field (Mattila \cite{mattila74}; FitzGerald et~al.\
\cite{fitzgerald76}). The bright rim maximum occurs at a radius
corresponding to extinction A$_{\lambda} \approx 1.5^{\mathrm
m}-2^{\mathrm m}$.

Laureijs et~al.\ (\cite{laureijs91}) observed the 60\,$\mu$m emission
to decline relative to 100\,$\mu$m emission in a narrow transition
layer around the cloud edge. They explained this behaviour by assuming
a separate grain component at 60\,$\mu$m that undergoes a modification
of properties in a transition layer.

Laureijs et~al.\ (\cite{laureijs95}) have used IRAS images, CO
molecular line observations and blue extinction values from star
counts for a large-scale study of the L~134 cloud complex, including
L~134, L~169 and L~183. Using $^{13}$CO observations, they found 18
clumps in the complex and derived their properties. The clumps follow
clear size vs.\ linewidth and luminosity vs.\ size relationships.  An
analysis and discussion of the dust and molecular gas properties in
L~183 based on the ISOPHOT 100 and 200\,$\mu$m emission maps has
recently been presented by Juvela et~al.\ (\cite{juvela02}).

L~183 has been a favourite target for molecular line observations
which have revealed it as a rich source of molecular species (see
e.g.\ Swade \cite{swade89a}, \cite{swade89b}; Dickens et~al.\
\cite{dickens}).  Lee et~al.\ (\cite{lee01}) have classified L~183 as
a strong infall candidate.  The optically thick CS($J$=2-1) lines in
the inner region of the cloud show a double peak with the blue
component brighter than the red one, which is characteristic of inward
motions. Similar profiles suggesting infall motions have been observed
for HCO$^+$($J$=3--2) (Gregersen \& Evans \cite{gregersen00}) and for
CS($J$=2--1) by Snell et~al.\ (\cite{snell82}).

Mapping of the millimetre and submillimetre continuum dust emission by
Ward-Thompson et~al.\ (\cite{wthompson94}, \cite{wthompson99}) has
revealed a small extended core in the center of L~183. This source has
been detected at 800, 1100 and 1300\,$\mu$m, and has FWHM dimensions
of $60\arcsec \times 40\arcsec$ (0.032 $\times$ 0.021\,pc) at
800\,$\mu$m. According to Ward-Thompson et~al.\ (\cite{wthompson94}),
the core is probably pre-protostellar in nature, i.e.\ at an earlier
stage than an accreting Class~0 protostar (for the definition of a
Class~0 object see Andr\'e et~al.\ \cite{andre93}).  The submillimetre
emission from the core is consistent with the dust being heated
externally by the general interstellar radiation field.  Recent
observations by Ward-Thompson et~al.\ (\cite{wthompson00}) have
revealed that the continuum emission extends further to the south, and
that there is another emission maximum located at a position about
1.5$\arcmin$ south of the previously detected maximum, with a FWHM
size of $120\arcsec \times 60\arcsec$. From sub-mm polarization
observations Ward-Thompson et~al.\ (\cite{wthompson00}) measured a
magnetic field direction which is at an angle of $34\degr \pm 6\degr$
to the minor axis of the core, and found evidence for decreasing
polarization at the highest continuum emission intensities.

Ward-Thompson et~al.\ (\cite{wthompson02a}) have mapped the core of
L~183 at 90, 170 and 200\,$\mu$m with ISOPHOT. The core was not
detected at 90\,$\mu$m wavelength. The dust temperature derived from
170 and 200\,$\mu$m data showed no temperature gradient across the
core.

In this study, we search for pre- and protostellar objects in the
cloud. Our maps are much bigger than previous (sub)millimetre
continuum maps.  The derived fluxes of the sources between
120--200\,$\mu$m, located near the peak or in the Wien regime of the
assumed blackbody radiation, are essential for temperature and thus
mass determination of the objects.  Only if the masses of the objects
are determined can we study their dynamical state.  Furthermore, we
study the relation of young (proto)stellar objects with the properties
of the underlying dust in the cloud such as temperature and column
density.

In the context of L~183, the term 'pre-protostellar core' has been
previously used to refer to the two known (sub)mm continuum sources
within the cloud (Ward-Thompson et~al.\ \cite{wthompson94},
\cite{wthompson00}), and also to the cloud core itself in which these
sources are embedded (Ward-Thompson et~al.\ \cite{wthompson02a}).
Throughout this article, we use the term 'source' to refer to objects
in L~183 which are unresolved in our maps, such as the
previously-known continuum sources. By the term 'core', we refer to
the visually opaque cloud core which is larger than the sources within
it and resolved by ISO observations.

\section{Observations and data reduction}

\begin{figure*}
\caption{Optical red Digitized Sky Survey (DSS) ({\bf a,b}),
200\,$\mu$m ({\bf c}) and 100\,$\mu$m ({\bf d}) images of L~183. The
four unresolved sources are marked as blue dots; for identification
see Fig.~\ref{cmaps}.
{\bf a)} Digitized Sky Survey red plate. The image is scaled to best
show the opaque core of the cloud.
{\bf b)} Digitized Sky Survey red plate. The image is scaled to best
show details of faint surface brightness.
{\bf c)} False-colour image of 200\,$\mu$m surface brightness.  The
colour scale is in units of MJy\,sr$^{-1}$.
{\bf d)} False-colour image of 100\,$\mu$m surface brightness.
Zodiacal light of 4.0\,MJy\,sr$^{-1}$ has been subtracted.  The colour
scale is in units of MJy\,sr$^{-1}$}
\label{optIR}
\end{figure*}

\begin{table}
\caption[]{The properties of filters. $\lambda_{\mathrm{ref}}$ is the
reference wavelength, $\Delta \lambda$ is the width, and 
d$_{\mathrm{Airy}}$ is the diameter of the Airy disk}
\begin{flushleft}
\begin{tabular}{rrrrr}
\hline\noalign{\smallskip}
Filter  &  $\lambda_{\mathrm{ref}}$  &  $\Delta \lambda$  &  
d$_{\mathrm{Airy}}$  &  Pixel size					\\
  &  [$\mu$m]  &  [$\mu$m]  &  [$\arcsec$]  &  [$\arcsec$]  		\\
\noalign{\smallskip}
\hline\noalign{\smallskip}
C\_100 & 100 & 43.6 & 83.9 & 43.5$\times$43.5		\\
C\_120 & 120 & 47.3 & 101  & 89.4$\times$89.4		\\
C\_135 & 150 & 82.5 & 113  & 89.4$\times$89.4		\\
C\_200 & 200 & 67.3 & 168  & 89.4$\times$89.4		\\
\noalign{\smallskip}
\hline
\end{tabular}
\end{flushleft}
\label{obspecs}
\end{table}

\begin{table*}
\caption[]{The parameters of the observations. TDT is the Target
Dedicated Time number of the observation, P.A.\ is the position angle
measured from North to East. All observations were made on January
19th 1997}
\begin{flushleft}
\begin{tabular}{rrrrrrr}
\hline\noalign{\smallskip}
Filter  &  TDT	&  Raster steps  &  Map size  &  
Grid  &  P.A.   &  Remarks	         		\\
  &  &  &  [$\arcmin$]  &  [$\arcsec$]  &  [\degr]  & 	\\
\noalign{\smallskip}
\hline\noalign{\smallskip}
C\_100  &  43001032  &  14$\times$20  &  31.5$\times$30.8  &  135, 90  &   
  21.1  &  Map         \\
C\_200  &  43001031  &  10$\times$10  &  30.0$\times$30.0  &  180, 180 &   
  21.1  &  Map         \\
C\_100  &  43000411  &  13$\times$2   &  39$\times$3  &  180, 90  & 
 179.0  &  N-S stripe  \\
C\_120  &  43000413  &  10$\times$2   &  30$\times$3  &  180, 90  & 
 179.0  &  N-S stripe  \\
C\_135  &  43000412  &  10$\times$2   &  30$\times$3  &  180, 90  & 
 179.0  &  N-S stripe  \\
C\_200  &  43000414  &  10$\times$2   &  30$\times$3  &  180, 90  & 
 179.0  &  N-S stripe  \\
\noalign{\smallskip}
\hline
\end{tabular}
\end{flushleft}
\label{obsdata}
\end{table*}

The observations were made with the ISOPHOT instrument aboard the ISO
satellite, using the C100 and C200 detectors (Lemke et~al.\
\cite{lemke96}).  We have mapped $\sim 30 \times 30$ arc~min$^2$ of
the cloud at 100 and 200\,$\mu$m.  In addition, we have performed
linear north-south raster scans at 100, 120, 150 and 200\,$\mu$m
through the cloud core with a two way path, the return path being
shifted by 90$\arcsec$ to the west (see Fig.~\ref{cmaps} for the
raster scan positions and the orientation of the detector array).  The
raster scans were designed to cross the continuum source FIR~1.  All
observations were made using the observing template PHT22 in raster
mode.  Details of the detector/filter combinations and observational
parameters are given in Tables~\ref{obspecs} and \ref{obsdata}.

The data analysis was done using PIA\footnote{The ISOPHOT data
presented in this paper were reduced using PIA, which is a joint
development by the ESA Astrophysics Division and the ISOPHOT
Consortium (the ISOPHOT Consortium is led by the Max-Planck-Institute
for Astronomy (MPIA), Heidelberg). Contributing ISOPHOT Consortium
institutes are DIAS, RAL, AIP, MPIK, and MPIA} (ISOPHOT Interactive
Analysis) V~8.1 (Gabriel et~al.\ \cite{gabriel}). At the first
processing level, the detector ramps were corrected for non-linearity
of the detector response, glitches in ramps were removed by using the
two-threshold glitch recognition method, and the ramps were fitted
with 1st order polynomials. At subsequent levels the signals were
deglitched, reset interval correction was applied, signals were
linearized for the dependence of detector response on illumination,
and orbital position-dependent dark currents were subtracted.

For the flat-field correction, a statistical method was applied: the
value of the reference pixel at each raster position was correlated
with other pixels. Instead of comparing the reference pixel only with
pixels located at the same raster position, a mean of two pixels
located symmetrically around the reference pixel wass taken.  In this
way, it is possible to reduce the scatter caused by surface brightness
gradients in the image.

The primary intensity calibrators are the fine calibration source
(FCS) measurements, which bracket the actual map or raster scan
measurements. The C100 detector suffers from response transients
during the relatively short measurement time of the FCS. We have
corrected for this effect by modelling the signal transients with the
PIA signal drift interface, using the offset exponential as the
fitting function.  The C200 detector reacts faster, thus there is no
such transient behaviour.

We have compared our 100\,$\mu$m surface brightness calibration with
COBE/DIRBE data. To match our data to the much coarser DIRBE
resolution (0.7\degr$\times$0.7\degr), we have used the IRAS/ISSA
100\,$\mu$m map of the L~183 area as an intermediate step (for the
method see Lehtinen et~al.\ \cite{lehtinen01}). We have thus found a
scaling factor $I\mathrm{_{100\,\mu m}(ISOPHOT)}/I\mathrm{_{100\,\mu
m}(DIRBE)}=1.25$. Although this value is well within the combined
uncertainty of the absolute accuracies cited for ISOPHOT (20\%, see
Klaas et~al.\ \cite{klaas00}) and for DIRBE (11\%, see Hauser et~al.\
\cite{hauser98}) it is larger than the scaling factor obtained for a
sample of five mappings of different clouds where the ISOPHOT and
DIRBE 100\,$\mu$m surface brightnesses agree to within 15\% (Lehtinen
et al.\ \cite{lehtinen01}). We have thus rescaled our present ISOPHOT
100\,$\mu$m map dividing it by 1.25 to correspond to the DIRBE
calibration. The North-South raster scan data at 100\,$\mu$m are
scaled to the map.

A direct ISOPHOT vs.\ DIRBE comparison was done at 200\,$\mu$m by
averaging the ISOPHOT surface brightness over our whole map area and
comparing it with the nearest DIRBE pixel value. The resulting scaling
factor was 0.90, i.e.\ well within the uncertainties of each
instrument. No rescaling was applied to our ISOPHOT data at
200\,$\mu$m, nor at 120 and 150\,$\mu$m.

\section{Results}

\begin{figure}
\resizebox{\hsize}{!}{\includegraphics{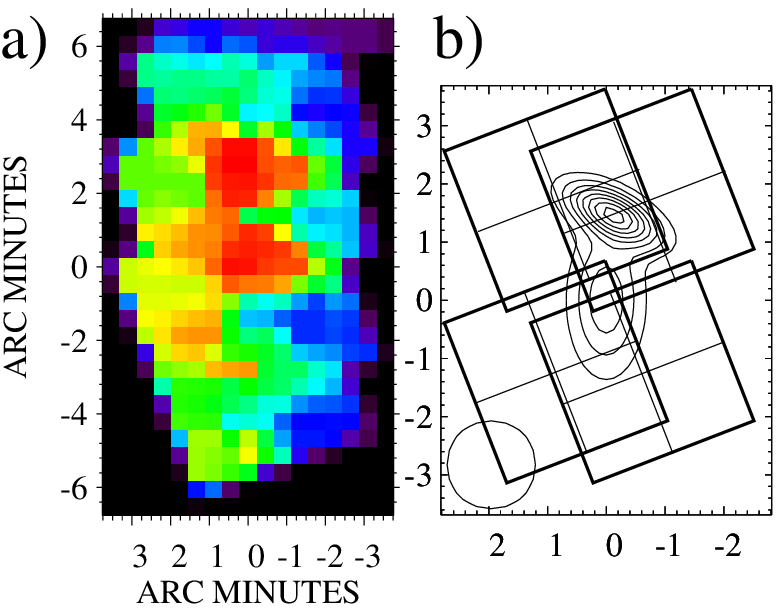}}
\caption{ ({\bf a}) Map made from the 200\,$\mu$m stripe scan data.
The pixel size is 30\arcsec. We note that this map is just to
illustrate that the two continuum sources detected by Ward-Thompson
et~al.\ (\cite{wthompson94}, \cite{wthompson00}) can be detected as
separate sources in the stripe scan data.  The details in this map are
smaller than the actual resolution of the stripe scan data.
({\bf b}) The model at 200\,$\mu$m used to derive the fluxes of FIR~1
and FIR~2. The positions and sizes of the sources are those given by
Ward-Thompson et~al.\ (\cite{wthompson94}, \cite{wthompson00}).  The
inclined boxes show the locations of the 2$\times$2 pixel C200
detector array in the North-South stripe scan.  The contours of the
continuum sources go from 0.8 to 6.4 in steps of
0.8\,Jy\,pixel$^{-1}$. The FWHM of the ISO beam at 200\,$\mu$m is
shown at lower left hand corner. The center position in both maps is
the position of FIR~2 (see Table~\ref{sources}) }
\label{model}
\end{figure}

\begin{figure*}
\resizebox{\hsize}{!}{\includegraphics{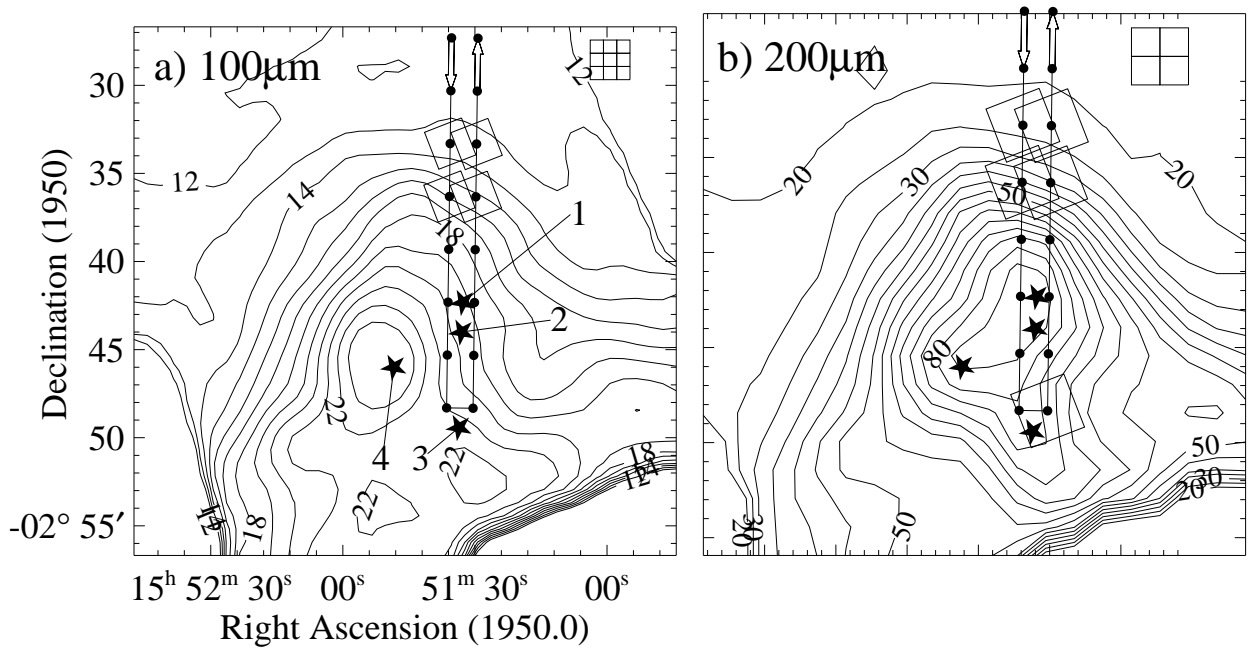}}
\caption{Contour maps of 100\,$\mu$m ({\bf{a}}) and 200\,$\mu$m
({\bf{b}}) surface brightness.  The detected unresolved sources are
marked with asterisks. The additional raster scan positions
(center of the detector array) observed at 100, 120, 150 and
200\,$\mu$m are indicated by dots.  The orientation of the C100 and
C200 detector is shown for four raster scan positions. The sizes of
the C100 and C200 detector pixels are shown at the upper right
corners.
{\bf{a)}} The contours are from 12 to 23\,MJy\,sr$^{-1}$ in steps of
1\,MJy\,sr$^{-1}$.  Zodiacal light of 4.0\,MJy\,sr$^{-1}$ has been
subtracted.
{\bf{b)}} The contours are from 20 to 85\,MJy\,sr$^{-1}$ in steps of
5\,MJy\,sr$^{-1}$.  The chance location of FIR~3 within the detector
area is shown}
\label{cmaps}
\end{figure*}

\begin{figure}
\resizebox{\hsize}{!}{\includegraphics{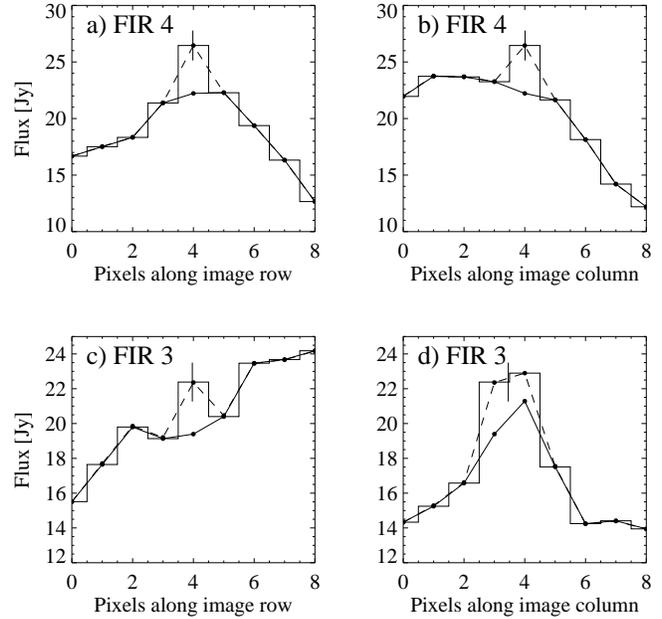}}
\caption{Profiles along image rows and columns through the sources
FIR~3 and 4 at 200\,$\mu$m.  The solid line is a fit to the background,
the dotted line shows the fitted point source profile. The vertical line
shows the fitted position of the source}
\label{profiles}
\end{figure}

\begin{figure*}
\resizebox{\hsize}{!}{\includegraphics{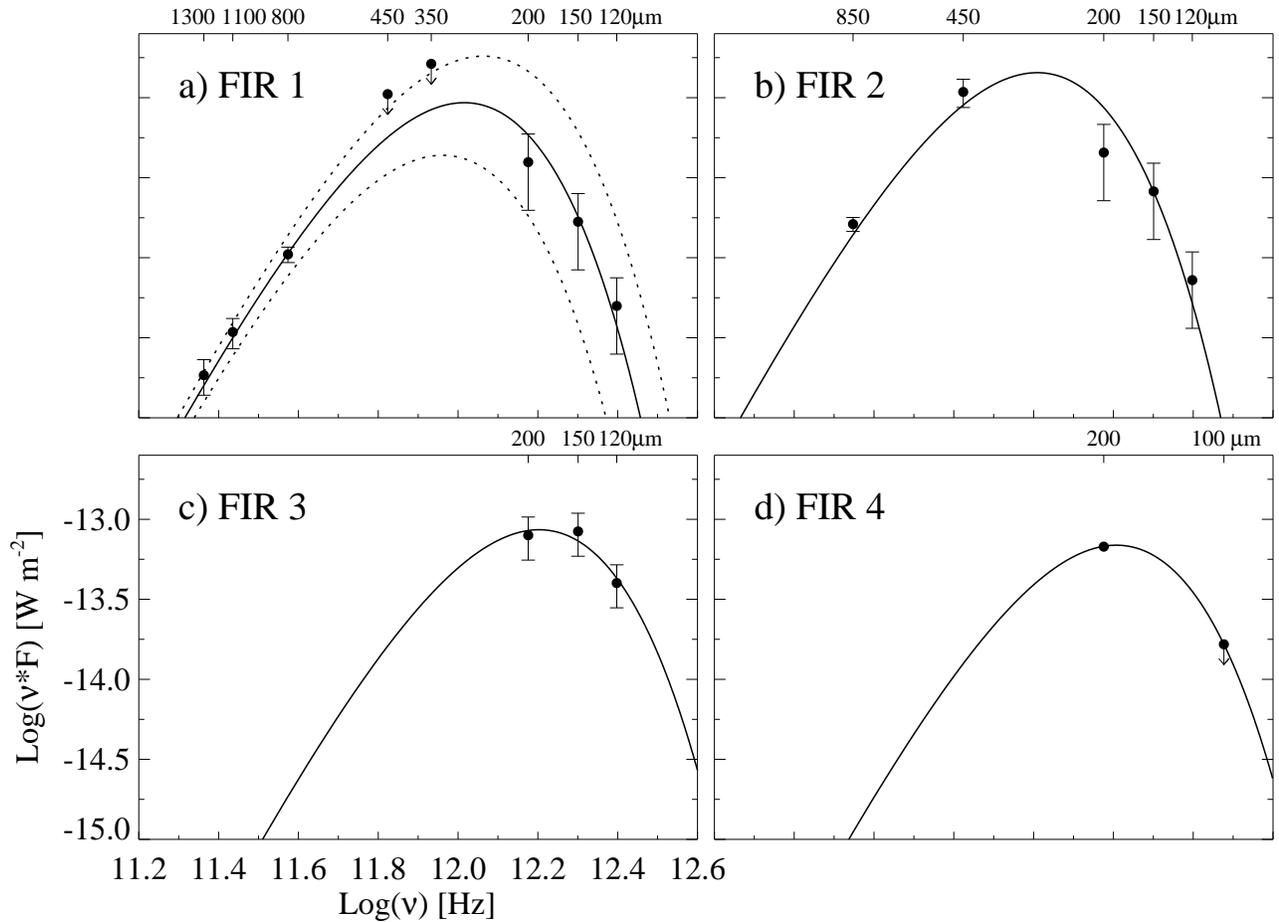}}
\caption{The spectral energy distributions of the unresolved sources.
The 100, 120, 150 and 200\,$\mu$m flux density values are our ISOPHOT
measurements, others are from the literature.  All fluxes are total
fluxes (see Sect. 3.2).  The solid lines are modified blackbody fits
($\beta=2$) to the data points.
{\bf{a)}} FIR~1. Derived temperature T=8.4\,K.  The dotted lines are
blackbodies which have 1\,K higher and lower temperatures than the
solid line, demonstrating the accuracy of the derived temperature.
{\bf{b)}} FIR~2. Derived temperature T=8.2\,K.
{\bf{c)}} FIR~3. Derived temperature T=12.8\,K.
{\bf{d)}} FIR~4. The solid line is a modified blackbody curve
(T=12.9\,K), with a temperature equal to the colour temperature
derived from the 100 and 200\,$\mu$m flux densities}
\label{seds}
\end{figure*}

\begin{table*}
\caption[]{Positions, sizes, and non colour-corrected total flux
densities of unresolved sources in L~183.  The 1-$\sigma$ error values
for the flux densities include an estimated 50\% error for FIR~1 and 2
and 30\% error for FIR~3 and 4}
\begin{flushleft}
\begin{tabular}{rrrrrrrrr}
\hline\noalign{\smallskip}
Name & FWHM size & \multicolumn{2}{c}{Position} & F$_{\nu}$(100\,$\mu$m) & 
F$_{\nu}$(120\,$\mu$m) & F$_{\nu}$(150\,$\mu$m) & 
F$_{\nu}$(200\,$\mu$m) & Remarks 					     \\
 & [\arcsec] & $\alpha$(1950) & $\delta$(1950) & [Jy] & [Jy] & [Jy] & [Jy] & \\
\noalign{\smallskip}
\hline\noalign{\smallskip}
FIR~1 & 60$\times$40 & $15^{\rm h} 51^{\rm m} 32\fs7$ 	
      & $-2\degr 42\arcmin 19\arcsec$ 
      & - & 0.14$\pm$0.07 & 0.76$\pm$0.38 & 2.6$\pm$1.3 & 1 		\\
FIR~2 & 120$\times$60 & $15^{\rm h} 51^{\rm m} 33\fs0$ 	
      & $-2\degr 44\arcmin 00\arcsec$ 
      & - & 0.59$\pm$0.30 & 2.3$\pm$1.2 & 5.2$\pm$2.6 & 1 		\\
FIR~3 & & $15^{\rm h} 51^{\rm m} 33\fs6$ 			
      & $-2\degr 49\arcmin  25\arcsec$ 
      & - & 1.6$\pm$0.5 & 4.2$\pm$1.4 & 5.3$\pm$1.8 & 			\\
FIR~4 & & $15^{\rm h} 51^{\rm m} 48\fs4$ 			   
      & $-2\degr 46\arcmin 00\arcsec$ 
      & $<$0.6 & - & - & 4.7$\pm$1.6 &  				\\
\noalign{\smallskip}
\hline
\end{tabular}
\end{flushleft}
$^1$Sizes and positions from Ward-Thompson et~al.\ (\cite{wthompson94}, 
                                          \cite{wthompson00})
\label{sources}
\end{table*}       

\begin{table*}
\caption[]{The parameters for unresolved sources.  T$_{\mathrm d}$ is
the dust temperature.  $\lambda_0$ is the wavelength (corresponding to
$\nu_0$) where the optical depth is unity.  The sub-mm luminosity
L$_{\mathrm{sub-mm}}$ is the luminosity of the fitted blackbody
function, integrated longward of 350\,$\mu$m. L$_{\mathrm{bol}}$ is
the bolometric luminosity, N(H$_2$) is the mean molecular hydrogen
column density, and n(H$_2$) is the mean number density assuming a
constant density sphere with size equal to the geometric mean of the
minor- and major-axis FWHM sizes. M is the total (gas plus dust) mass,
and M$_{\mathrm{vir}}$ is the virial mass.  The quoted errors are
1-$\sigma$ errors based on Monte Carlo error estimation, including the
estimated errors of fluxes only.  Note that the parameters of FIR~3
have been calculated by assuming that its size is equal to the mean
value of the sizes of FIR~1 and 2}
\begin{flushleft}
\begin{tabular}{rrrrrrrrrr}
\hline\noalign{\smallskip}
Name  				&
T$_{\mathrm d}$ 		&  $\lambda_0$ 			&
L$_{\mathrm{sub-mm}}$   	&  L$_{\mathrm{bol}}$   	&  
$\frac{L_{\mathrm{sub-mm}}}{L_{\mathrm{bol}}}$			& 
N(H$_2$)  & 	n(H$_2$)        &  M  & M$_{\mathrm{vir}}$ 	\\

                                &
[K]				&  [$\mu$m]			&
[L$_{\sun}$]			&  [L$_{\sun}$]			&
				&  [10$^{22}$\,cm$^{-2}$]	&
[10$^6$\,cm$^{-3}$]   &   [M$_{\sun}$]	&  [M$_{\sun}$]  	\\
\noalign{\smallskip}
\hline\noalign{\smallskip}
FIR~1  &  8.4$\pm 0.4$ &  63$\pm 6$ & 0.014 & 0.036 & 0.4 & 14
       & 2.6 & 1.4 & 0.4                 	\\
FIR~2  &  8.2$\pm 0.4$ &  48$\pm 8$ & 0.022 & 0.056 & 0.4 & 8.1
       & 0.9 & 2.4 & 0.7                 	\\
FIR~3  & 12.8$\pm 1.5$ &  12$\pm 5$ & 0.004 & 0.034 & 0.1 & 0.4
       & 0.05 &  0.2                  	\\
FIR~4  & $<$12.9       &              &       & $<$0.05 &   &  
&   &  -		 	        \\
\noalign{\smallskip}
\hline
\end{tabular}
\end{flushleft}
\label{fitparameters}
\end{table*}

\subsection{Unresolved sources in L~183}

The ISOPHOT far-IR maps at 100\,$\mu$m and 200\,$\mu$m, together with
optical images, are shown in Fig.~\ref{optIR}. We have searched for
point-like objects by visual inspection.  Neither one of the sub-mm
continuum sources detected by Ward-Thompson et~al.\
(\cite{wthompson94}, \cite{wthompson99}, \cite{wthompson00}) is
detected in the 100\,$\mu$m map (see Fig.~\ref{optIR}d).  At
200\,$\mu$m there is an emission maximum at the position of the
sources (see Fig.~\ref{optIR}c).  They have an angular distance of
$\sim$1 pixel, so they are not detected as separate sources in the
200\,$\mu$m map.  However, in the C200 North-South stripe scans the
resolution is better due to overlapping pixels (see
Fig.~\ref{model}b).  A map made from the 200\,$\mu$m stripe scan data
clearly shows two emission maxima whose positions correspond to
positions of the continuum sources of Ward-Thompson et~al.\ (see
Fig.~\ref{model}a).  In order to derive the total fluxes of these
sources, our sources FIR~1 and 2, we have modelled them by two-axial
gaussian surfaces having position angles and FWHM sizes as given by
Ward-Thompson et~al.\ (\cite{wthompson94}, \cite{wthompson00}).  We
have not made a 2-D map of the stripe scan data, but have treated them
as a 1-D time-ordered data for each pixel.  The background has been
determined by fitting the 1-D data with cubic splines, excluding the
positions where the sources contribute to the flux. The flux seen by
each pixel has been calculated by using our 2-D model and ISO beam
profile. The fluxes of the sources have been determined by minimizing
the difference between the modelled and observed fluxes in a least
squares sense.  The model is shown in Fig.~\ref{model}b, and the
derived total fluxes are given in Table~\ref{sources}.  The accuracy
of the derived fluxes is mainly limited by our ability to subtract the
strongly-varying background at the edge of the cloud. We therefore
give an estimated error of $\pm 50$\,\% for the fluxes.

A visual inspection of the 200\,$\mu$m map further reveals two
previously unknown unresolved sources in the cloud, sources FIR~3 and
4 (see Fig.~\ref{optIR}c).  Neither of them is detected in the
100\,$\mu$m map.  FIR~3 is by chance located within the C200 detector
array at one stripe scan position (see Fig.~\ref{cmaps}\,b), which
enables us to also estimate its flux at 120 and 150\,$\mu$m. We take
the flux of the pixel closest to the source, subtract from that the
mean of the other three pixels, and correct the flux by the point
spread function fraction for a point source centered on a pixel
(Laureijs \cite{laureijs99}).  An error of 30\% has been estimated for
the fluxes.  The 5.3\,Jy 200\,$\mu$m total flux derived this way is in
good agreement with the 5.8\,Jy flux derived from the map. Profiles
through FIR~3 along image rows and columns in the 200\,$\mu$m map are
shown in Fig.~\ref{profiles}c and d, respectively.

Source FIR~4 is a prominent unresolved source on the 200\,$\mu$m map
(Figs.~\ref{optIR}c and \ref{profiles}).  On the 100\,$\mu$m map there
is an extended brightness enhancement around the position of FIR~4
(Fig.~\ref{optIR}\,d), where the 100\,$\mu$m brightness reaches its
maximum. In order to derive the 100\,$\mu$m flux density, we have
first convolved the data to the 200\,$\mu$m resolution. Since it is
difficult to separate the source from the structured background, we
consider the derived flux density as an upper limit.  Profiles through
FIR~4 along image rows and columns in the 200\,$\mu$m map are shown in
Fig.~\ref{profiles}a and b, respectively.

We have searched in the IRAS Point Source (PSC) and Faint Source
Catalogs (FSC) for sources within our maps. We find two PSC sources,
IRAS 15522--0258 and 15523--0251, and one FSC source, F15507-0245.
None of them are detected by us. The IRAS sources are located outside
the cloud centre.  The PSC sources only have 100\,$\mu$m IRAS
fluxes. In the ISO maps, there is no enhancement of 100\,$\mu$m
surface brightness at the position of the PSC sources. As the ISO and
IRAS images at 100\,$\mu$m appear morphologically similar, it seems
that the two 100\,$\mu$m PSC sources are artefacts, and are a result
of IRAS scanning direction with respect to structures in the cloud.
The FSC source has a 12\,$\mu$m detection only and is thus not
expected to be detected at 100\,$\mu$m.

\subsection{The spectral energy distributions}

At $\lambda \ga 100\,\mu$m, the emission arises from 'classical' large 
grains, which are at an equilibrium temperature within the radiation
field.  We have thus fitted the flux densities at these wavelengths
with a modified blackbody of the form
\begin{equation}
  F_{\nu} = B_{\nu}(T_{\rm d})\,(1-\exp(-\tau_{\nu}))\,\Omega_{\rm s}
\end{equation}
where $B_{\nu}(T_{\mathrm d})$ is the Planck function at the dust
temperature $T_{\mathrm d}$, $\tau_{\nu}$ is the optical depth which
is assumed to vary with frequency as $\tau_{\nu} =
(\nu/\nu_{0})^{\beta}$, $\Omega_{\mathrm s}$ is the solid angle of the
emitting region, and $\nu_{0}$ is the frequency at which the optical
depth is unity.  We have assumed that the emissivity index ${\beta}$
is 2. For the source size $\Omega_{\mathrm s}$ of FIR~1 and 2 we
have used the FWHM sizes given by Ward-Thompson et~al.\
(\cite{wthompson94}; \cite{wthompson00}). The size of FIR~3 is unknown,
so we have assumed a size which is a mean value of the sizes of FIR~1
and 2.

The 100 to 200\,$\mu$m total fluxes measured with ISO are complemented
by values at longer wavelengths from Ward-Thompson et~al.\
(\cite{wthompson94}; \cite{wthompson99}) for FIR~1, and from
Ward-Thompson et~al.\ (\cite{wthompson02b}) for FIR~2.  Ward-Thompson
et~al.\ (\cite{wthompson94}) give the total flux of FIR~1 only at
800\,$\mu$m. From the given total flux and flux within a 18\arcsec
beam we derive their ratio of 10.4.  Because the other fluxes at 450
and 1100\,$\mu$m are given for the same 18\arcsec beam as the
800\,$\mu$m flux, we can use the ratio 10.4 to derive total fluxes.
This way we derive total fluxes of $<$15.6 and 1.1\,Jy at 450 and
1100\,$\mu$m, respectively.  Ward-Thompson et~al.\
(\cite{wthompson99}) have given for FIR~1 at 1.3\,mm a flux of 40\,mJy
within 13\arcsec beam. Thus the total flux is $(18/13)^2 \times 10.4
\times 40$\,mJy=0.8\,Jy.  The spectral energy distributions and fitted
modified blackbody functions for FIR~1, 2, 3 and 4 are shown in
Fig.~\ref{seds}.  The resulting temperatures are listed together with
other parameters in Table~\ref{fitparameters}. The uncertainties of
the fitted parameters have been estimated by Monte Carlo methods.

For FIR~1, 2 and 3 the ISO flux densities, determined near the peak of
the SEDs, accurately set the temperatures of the fitted blackbodies,
as demonstrated in Fig.~\ref{seds}.  The fitted temperatures are 8.4,
8.2 and 12.8\,K, respectively. The derived colour temperature of FIR~4
has an upper limit of about 13\,K. We note that the use of a single
temperature is a simplification, and that the source sizes may be
different at different wavelengths.  Ward-Thompson et~al.\
(\cite{wthompson02a}) derived a temperature of 10$\pm 3$\,K within a
150\arcsec circular core area of the cloud, which is within errors
equal to our value, $8.3 \pm 0.4$\,K.

\subsection{The luminosities, column densities, number densities, and masses 
of unresolved sources}

The bolometric luminosities, L$_{\mathrm{bol}}$, have been derived for
all sources by integrating over the fitted blackbody curves
(Table~\ref{fitparameters}).  For the bolometric luminosity of the
core of the cloud Ward-Thompson et~al.\ (\cite{wthompson02a}) derived
0.15\,L$_{\sun}$ by integrating the fluxes at 170\,$\mu$m and
200\,$\mu$m in a circular 150\arcsec aperture, combining these fluxes
with mm/submm fluxes, and fitting the SED with a modified blackbody.
The combined luminosity of IRS~1 and 2 derived by us,
0.09\,L$_{\sun}$, is in excellent agreement with the core luminosity
of Ward-Thompson et~al.\ when taking into account the different
distances used by us (110\,pc) and Ward-Thompson et~al.\ (150\,pc).

The mean column densities through FIR~1 and 2 have been derived from the
equation
\begin{equation}
N({\mathrm H}_2) = F_{\nu} / \left( \Omega_{\mathrm s}\, \mu\: m_{\mathrm H}\, 
                  \kappa_{\nu}\, B_{\nu}(T_{\mathrm d}) \right)
\end{equation}
where $F_{\nu}$ is the observed flux, $\mu=2.33$ is the mean molecular
weight, $m_{\mathrm H}$ is the mass of atomic hydrogen, and
$\kappa_{\nu}$ the dust opacity per unit mass (gas+dust) column
density (see e.g.\ Hildebrand \cite{hildebrand}; Chini et~al.\
\cite{chini87}). We have adopted $\kappa_{1300\,\mu
\rm{m}}=0.005$\,cm$^2$\,g$^{-1}$ which is valid for pre-protostellar
dense clumps (Motte et~al.\ \cite{motte98} and references therein),
assuming a gas-to-dust ratio of 100.  The value of
$\kappa_{1300\,\mu\rm{m}}$, and correspondingly the column density, is
uncertain by a factor of $\sim$2 (see e.g.\ Motte et~al.\
\cite{motte98}).  We obtain the values $N({\mathrm H_2}) \approx
14\,10^{22}$\,cm$^{-2}$ and 8.1\,$10^{22}$cm$^{-2}$ for FIR~1 and 2,
respectively.  These values are similar to the values in cloud cores
found by Motte et~al.\ (\cite{motte98}) in $\rho$~Ophiuchi cloud based
on millimeter continuum mapping, or by Bacmann et~al.\
(\cite{bacmann00}) based on ISOCAM absorption survey.  The difference
by a factor of about two in the column density of IRS~1 between us and
Ward-Thompson et~al.\ (\cite{wthompson99}) is due to the higher
temperature (12.5\,K) adopted by Ward-Thompson et~al.

Mean number densities $n$(H$_2$), assuming a constant density, 
have been calculated using the formula 
$n({\mathrm H_2}) = (3/4) N({\mathrm H_2}) / R$, where the cloud
radius $R$ for FIR~1 and 2 is assumed to be the geometric mean of the
minor- and major-axis FWHM sizes.

We are able to derive improved mass estimates because ISOPHOT flux
values enable accurate determination of dust temperature.  In order to
estimate the total (gas plus dust) mass $M$ of matter associated with
dust we use the equation
\begin{equation}
  M = \frac{ F_{\nu} D^{2} } { \kappa_{\nu} B_{\nu}(T_{\rm d}) }
\end{equation}
with $D$ the distance.  The fitted values of $\lambda_0$, i.e.\ the
wavelength where $\tau_{\rm dust} = 1$ (see
Table~\ref{fitparameters}), suggest that for all sources the emission
is optically thin at wavelengths $\ga 100\,\mu$m.  For FIR~1, we have
used the 1300\,$\mu$m flux in Eq.~(2), giving $M \approx
1.4$\,M$_{\sun}$. For sources FIR~2 and 3, we are using the 850 and
200\,$\mu$m fluxes, respectively. The value of $\kappa_{\lambda}$ has
been scaled from $\kappa_{1300\,\mu \rm{m}}$ by using a $\lambda^{-2}$
dependence.  The derived masses of FIR~2 and 3 are 2.4 and
0.2\,M$_{\sun}$, respectively.

The relative mass of the sources can be derived without the knowledge
of $\kappa_{\lambda}$ or $D$. Only the dust temperature and flux at
any optically thin wavelength are required. We have derived the
relative mass between FIR~1 and 2 using the optically thin 1.3\,mm
fluxes from the blackbody fit. The derived mass ratio between FIR~1
and 2 is 0.6.

\subsection{Extended dust emission}

\begin{figure}
\resizebox{\hsize}{!}{\includegraphics{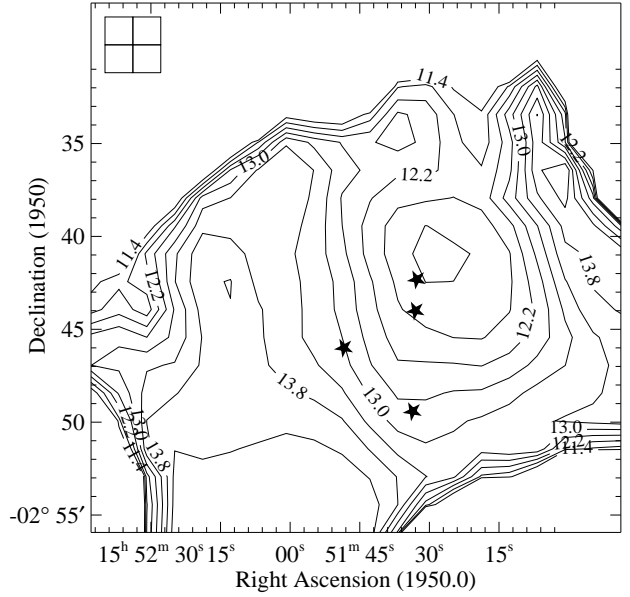}}
\caption{Dust temperature as derived from the 100 and 200\,$\mu$m
maps. The unresolved sources have been removed from the 200\,$\mu$m
map before temperature calculation.  The contours are from 11.4 to
14.2 in steps of 0.4\,K.  The detected sources are marked as stars.
The temperature is not determinable outside the cloud due to the
background subtraction.  The size of the C200 detector is shown at the
upper left corner}
\label{temperature}
\end{figure}

\begin{figure}
\resizebox{\hsize}{!}{\includegraphics{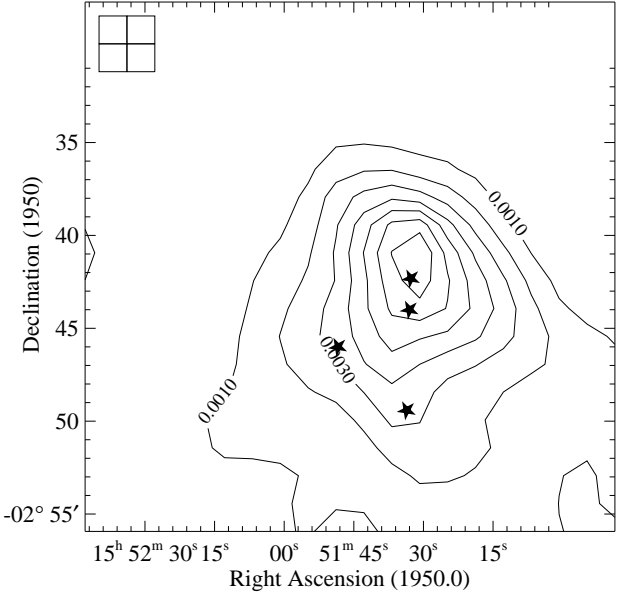}}
\caption{Map of beam-averaged optical depth at 200\,$\mu$m after
subtraction of the 4 point sources.  The contours are from 1$\times
10^{-3}$ to 7$\times 10^{-3}$ in steps of 1$\times 10^{-3}$}
\label{optdepth}
\end{figure}

A detailed analysis of the extended dust emission is presented by
Juvela et~al.\ (\cite{juvela02}). In this paper we will present the
derivation of the temperature and column density distribution of the
dust.

The deep red Schmidt image in Fig.~\ref{optIR} shows faint surface
brightness details across the core, which are probably structures in
front of the dense core and are seen as bright regions due to
scattered light.  In the optical image, the eastern part of the cloud
is covered by diffuse material.  The dense part of the cloud which is
delineated by the background stars extends untill $\alpha \approx
15^{\mathrm{h}}52^{\mathrm{m}}$.  Outside the diffuse material in the
northern and western parts of the cloud the optical edge of the cloud
closely follows the edge of 200\,$\mu$m surface brightness. The region
of strong ($\ga 80$\,MJy\,sr$^{-1}$) 200\,$\mu$m emission extends east
behind the diffuse emission and follows the opaque cloud boundary.
The diffuse emission is located in front of the cloud core.

There is a major difference between the 100 and 200\,$\mu$m maps: the
200\,$\mu$m maximum around the sources FIR~1 and 2 has no counterpart
at 100\,$\mu$m, i.e.\ there is a 200\,$\mu$m emission excess relative
to 100\,$\mu$m.  This could be due to a low dust temperature.  Other
explanations are that the dust emissivity at 200\,$\mu$m increases in
the dense core due to grain coagulation (Cambr\'esy et al.\
\cite{cambresy01}; Stepnik et~al.\ \cite{stepnik02}), or that the two
wavelengths trace different dust populations.

In order to derive the colour temperature of dust, the fluxes at 100
and 200\,$\mu$m have been fitted with a modified blackbody function of
the form
\begin{equation}
I_{\nu} \propto \nu^{\beta} B_{\nu}(T_{\mathrm d})
\end{equation}
where $\beta$ is the emissivity index and $T_{\mathrm d}$ is the dust
temperature.  We convolve the 100\,$\mu$m map to the 200\,$\mu$m
resolution, and subtract the background (determined at the upper right
corner in Fig.~\ref{cmaps}c and d).  While fitting the modified
blackbody, we iteratively make colour corrections until the difference
between subsequent temperatures is less than 0.1\,K.  The temperature
map (using $\beta=2$) is shown in Fig.\ref{temperature}.  The minimum
at the center of the cloud is $T_{\mathrm d} \approx$11.1\,K.

In the optically thin regime, the optical depth is given by observed
surface brightness
\begin{equation}
\tau_{\nu} = I_{\nu} / B_{\nu}(T_{\mathrm d})
\end{equation}
The map of beam-averaged optical depth at 200\,$\mu$m is shown in
Fig.\ref{optdepth}.  The optical depth sharply decreases north of the
sources FIR~1 and 2, while the decrease towards the south is smoother,
similar to the morphology of the 200\,$\mu$m map
(Fig.~\ref{cmaps}\,c).  The highest optical depth,
$\tau_{\mathrm{200\,\mu m}} \approx 7 \times 10^{-3}$, is ten times
larger than the optical depth through the center of the Thumbprint
Nebula (Lehtinen et~al.\ \cite{lehtinen98}), a globule without star
formation.

\section{Discussion}

\subsection{Are FIR~1 and 2 gravitationally bound objects ?}

\begin{table}
\caption[]{The derived energies for FIR~1 and 2 in units of
10$^{34}$\,J. The energy ratio is the ratio
$\mid \mathrm{E_{pot}}\mid/ (\mathrm{E_{therm}}+\mathrm{E_{turb}}-
\mathrm{E_{ext}}+0.5 \mathrm{E_{mag}})$
which has a value of 2 in the case of virial equilibrium.
The magnetic energy E$_{\mathrm{mag}}$ is based on estimated upper
limit for magnetic field strength $B<100\,\mu$G}
\begin{flushleft}
\begin{tabular}{rrrrrrr}
\hline\noalign{\smallskip}
Source  &  E$_{\mathrm{therm}}$  &  E$_{\mathrm{turb}}$  &  
E$_{\mathrm{ext}}$  &  E$_{\mathrm{mag}}$  &  
$\mid$E$_{\mathrm{pot}}\mid$  &  Energy            	\\
  &  &  &  &  &  &  ratio   				\\
\noalign{\smallskip}
\hline\noalign{\smallskip}
FIR~1  & 15 & 2.3 & 0.35 & $<1.0$  &  91 & 5.2 	 	\\
FIR~2  & 25 & 3.4 &  1.8 & $<5.4$  & 150 & 5.1 		\\
\noalign{\smallskip}
\hline
\end{tabular}
\end{flushleft}
\label{energies}
\end{table}

By definition, a pre-protostellar object has to be gravitationally
bound.  Thus, the key question concerning the far-IR/sub-mm sources in
L~183 is whether they are gravitationally bound, possibly already
collapsing objects, or whether they are just some random density
enhancements which will subsequently dissolve again.

The improved mass estimates of FIR~1 and 2 enable us to make estimates
of their dynamical state. A full treatment of virial equilibrium
requires the consideration of gravity ($E_{\mathrm{pot}}$) and surface
pressure ($E_{\mathrm{ext}}$) energies (compressing forces), and total
kinetic ($E_{\mathrm{kin}}$) and magnetic ($E_{\mathrm{mag}}$)
energies (supporting forces).  The condition for virial equilibrium is
then (McKee et~al.\ \cite{mckee93} and references therein)
\begin{equation}
2 (E_{\mathrm{kin}} - E_{\mathrm{ext}}) + E_{\mathrm{mag}} + 
E_{\mathrm{pot}} = 0
\end{equation}

For the kinetic energy we have included the thermal energy of
molecules and the turbulent energy. The thermal energy of molecules is
\begin{equation}
E_{\mathrm{therm}} = \frac{3}{2} \mathcal{N} k T
\end{equation}
where $\mathcal{N}$ is the total number of molecules, and $T$ is the
kinetic temperature, and $k$ is the Boltzmann constant.  The turbulent
energy is
\begin{equation}
E_{\mathrm{turb}} = \frac{3}{2} M \sigma_{\mathrm{turb}}^{2}
\end{equation}
where $M$ is the cloud mass, and $\sigma_{\mathrm{turb}}$ is the
three-dimensional non-thermal (turbulent) velocity dispersion of the
molecular gas, $\sigma_{\mathrm{turb}} = \sqrt{\Delta V^2/(8 \ln 2) -
k T / m}$, where $\Delta V$ is the observed linewidth (FWHM) and $m$
is the mass of the molecule in consideration.  We have estimated the
velocity dispersion from NH$_3$ observations which have $\Delta V =
0.28\,$km\,s$^{-1}$ (Ungerechts \cite{ungerechts80}).  For the kinetic
temperature, we have used a value of 10\,K. We note that these values
of velocity dispersion and kinetic temperature are values for the gas
in which FIR~1 and 2 are embedded in.

The external energy due to an external surface pressure acting on a
core is approximately (Vall\'ee \cite{vallee00})
\begin{equation}
E_{\mathrm{ext}} = ( \frac{ n_{\mathrm{ext}} }{ 400\,\mathrm{cm}^{-3} } )
                   ( \frac{ R }{ 0.6\,\mathrm{pc} } )^3
             ( \frac{ \sigma_{\mathrm{ext}} }{ 1.0\,\mathrm{km\,s}^{-1} } )^2
             \times 5 \times 10^{37}\, \mathrm{J}
\end{equation}

where $n_{\mathrm{ext}}$ and $\sigma_{\mathrm{ext}}$ are the density
and velocity dispersion of the gas in which a core is embedded, and
$R$ is the core radius. We have used the values $n_{\mathrm{ext}}=3.0
\times 10^4$\,cm$^{-3}$ (Swade \cite{swade89b}) and
$\sigma_{\mathrm{ext}}=0.3$\,km\,s$^{-1}$ as given by $^{13}$CO(2--1)
and C$^{18}$O(1--0) observations (Juvela et~al.\ \cite{juvela02}).
For $R$ we have used the HWHM radii; 0.013 and 0.023\,pc for FIR~1 and
2, respectively.

The total net magnetic energy is 
\begin{equation}
E_{\mathrm{mag}} = \frac{4 \pi R^3}{3} \frac{B^2}{8 \pi}
\end{equation}
which can be approximated by the formula (Schleuning 
\cite{schleuning98})
\begin{equation}
E_{\mathrm{mag}} = ( \frac{ B }{ 100\,\mu \mathrm{G} } )^2
                    ( \frac{ R }{ 0.6\,\mathrm{pc} } )^3
                    \times 10^{39}\, \mathrm{J}
\end{equation}
where $B$ is the magnetic field strength.  Recently, Crutcher \&
Troland (\cite{crutcher00}) have determined the line-of-sight magnetic
field strength of $\sim 11\,\mu$G in the L~1544 core using OH Zeeman
measurements with a 3\arcmin\ FWHM beam. Upper limits toward seven
other dense cores gave $B_{\mathrm{LOS}} \la 20-30\,\mu$G. Crutcher
(\cite{crutcher99}) has determined an upper limit
$B_{\mathrm{LOS}}<16\,\mu$G for L~183 with a much larger (18\arcmin)
beam. Allowing for another factor of four for the line-of-sight
projection effects and for the higher density in the L~183 FIR~1 and
FIR~2 cores we adopt $B <100\,\mu$G as an upper limit for the magnetic
field strenght.

The radial density profile $\rho(r)$ of FIR~1 has been estimated by
Ward-Thompson et~al.\ (\cite{wthompson94}) based on the observed
800\,$\mu$m continuum flux density profile, and assuming that the core
is isothermal; the density is $\rho(r) \propto r^{-1.25}$ from the
center to a distance of $\sim 28\arcsec$ and $\rho(r) \propto r^{-2}$
furtherout.  In a similar way, we have used Fig.~2 of Ward-Thompson et
al.\ (\cite{wthompson00}) to derive the radial flux density profile
for FIR~2.  We find that a single power law for space density is not
consistent with the observed radial flux density profile. Instead, two
power laws with the same power law exponents as in the case of FIR~1
fit the data well.  We have used these density profiles, with the
addition of a constant-density central core having a radius of
1/10\,th of the total radius. We have numerically calculated the
gravitational potential energy E$_{\mathrm{pot}}$ for FIR~1 and 2. For
the sizes of FIR~1 and 2, we have used the FWHM sizes given by
Ward-Thompson et~al.\ (\cite{wthompson94}, \cite{wthompson00}).

The calculated energies are shown in Table~\ref{energies}.  For both
sources we have 
$E_{\mathrm{therm}} > E_{\mathrm{turb}} > E_{\mathrm{ext}}$.  
The ratio $\mid \mathrm{E_{pot}}\mid/
(\mathrm{E_{therm}}+\mathrm{E_{turb}}- \mathrm{E_{ext}}+0.5
\mathrm{E_{mag}})$ has a value of two in virial equilibrium. For both
FIR~1 and 2, the ratio has a value of about 5.

Neglecting the magnetic energy and external pressure, the virial 
mass can be written in the form 
\begin{equation}
M_{\mathrm{vir}} = (\sigma^2 D) / (2 G)
\end{equation}
where $D$ is the cloud diameter, $G$ is the gravitational constant,
and $\sigma$ is the three-dimensional velocity dispersion for a mean 
gas particle mass, 
\begin{equation}
\sigma=\sqrt{ 3 \left( \frac{kT}{\overline{m}} + \left( 
       \frac{\Delta V^2}{8 \ln(2)} - \frac{kT}{m} \right) \right) }
\end{equation}
where $\overline{m}$ is the mean molecular mass, and $m$ is the mass
of the molecule used for observations.  This form of virial
equilibrium is exact for a polytrope of index $n=2$ (Larson
\cite{larson81}). Using the values $\Delta V = 0.28\,$km\,s$^{-1}$
from NH$_3$ observations and T=10\,K we obtain the virial masses 0.4
and 0.7\,M$_{\sun}$ for FIR~1 and 2, respectively, which are about one
third of the observed masses.  Taking into account all uncertainties
of the calculation, we conclude that the cores have masses which are
higher than or comparable to their virial masses, and are thus
gravitationally bound.

\subsection{Location of the sources relative to cold dust and gas}

The sources FIR~1 and 2 are located very close to the dust column
density maximum (see Fig.~\ref{optdepth}).  If we exclude the
possibility of a chance superposition, there is thus evidence that
these sources were born at the very center of L~183. At the positions
of FIR~3 and 4, the column density has decreased to about half of the
maximum value.

The NH$_3$ emission is considered to be a good tracer of star-forming
dense cores. The NH$_3$(1,1) line intensity maps of Ungerechts
(\cite{ungerechts80}) and Swade (\cite{swade89a}, \cite{swade89b})
show a north-south elongated structure with two maxima. The maxima are
separated by about 3 arcminutes, and the sources FIR~1 and 2 are
located near the peak of the southern maximum. FIR~3 is located at the
very southern tip of the NH$_3$ map of Swade (\cite{swade89a}). FIR~4
lies outside the existing NH$_3$ and many other high-density tracer
maps, but within the C$^{18}$O map of Juvela et~al.\
(\cite{juvela02}).

In the C$^{18}$O(1--0) channel maps of Juvela et~al.\
(\cite{juvela02}) FIR~3 coincides with a clump which is most clearly
separated at a velocity interval of 1.65--2.15\,km\,s$^{-1}$, while
emission around FIR~4 is strongest at velocities
2.15--2.9\,km\,s$^{-1}$. Emission around FIR~1 and 2 is most prominent
between 2.15--2.9\,km\,s$^{-1}$. At higher velocities, between
2.9--3.4\,km\,s$^{-1}$, the most intense emission is around FIR~1 and
2.  It is thus possible that FIR~3 is associated with a separate cloud
component which is isolated both in spatial and velocity space.

Juvela et~al.\ (\cite{juvela02}) present maps of H$^{13}$CO$^+$(1--0)
and DCO$^+$(2--1) emission. FIR~3 and 4 are located outside the mapped
areas. In particular the distribution of the cold gas tracer
DCO$^+$(2--1) looks very similar to the NH$_3$ distribution; it has
two maxima, the sources FIR~1 and 2 are located near the southern
maximum, and the emission continues south towards FIR~3. The extent
and structure of H$^{13}$CO$^+$(1--0) emission are rather similar to
DCO$^+$(2--1); FIR~1 and 2 are located near its maximum and the
emission continues towards FIR~3.

\subsection{Evidence for infall and outflow motions in L~183}

\begin{figure}
\resizebox{\hsize}{!}{\includegraphics{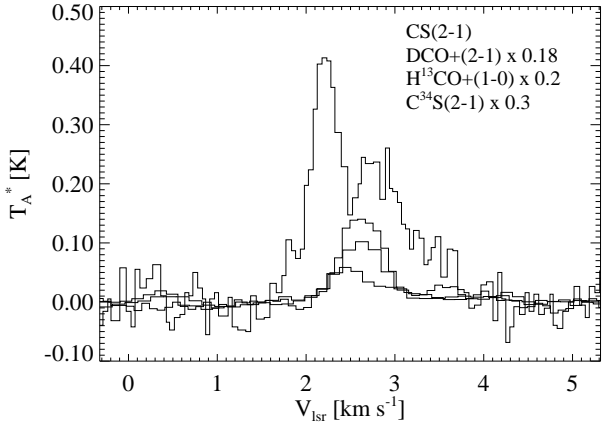}}
\caption{Molecular line profiles showing the infall asymmetry towards
FIR~1 and 2.  The CS(2--1) line is from Lee et~al.\ (\cite{lee99}),
others are from us}
\label{infall}
\end{figure}

The detection of an optically thick molecular line (such as CS(2--1),
HCO$^+$(3--2)) with redshifted self-absorption, possibly with an
absorption dip located at the same velocity as an optically thin line,
is considered to be a strong evidence of inward motions.  Recent
observations of starless dense cores have shown that extended inward
motions are common and may be a necessary step in the star-formation
process (Lee et~al.\ \cite{lee01}).  The molecular line maps of L~183
by Lee et~al.\ (\cite{lee01}) show CS(2--1) line profiles with infall
asymmetry over an extended area of some 5\arcmin--6\arcmin\, around
FIR~1 and 2, and N$_2$H$^+$(1--0) lines peaking at the absorption dip
of CS(2--1) lines. Lee et~al.\ (\cite{lee01}) have derived an infall
radius of $3.8\arcmin$, which is about the same as the combined
extension of FIR~1 and 2.

One might wonder if the large-scale infall is connected with the
build-up of the sources FIR~1 and 2, as they are the central sources
of the core.  Lee et~al.\ (\cite{lee01}) have estimated the infall
speeds and mass infall rates based on a simple two-layer radiative
transfer model for clouds having infall asymmetries in their line
profiles. They derive a mass infall rate for L~183 of $\sim 1-2 \times
10^{-5}$\,M$_{\sun}$\,yr$^{-1}$.  With this rate, the build-up of the
combined mass of FIR~1 and 2 would have taken $\sim 1.2 \times
10^5-2.3 \times 10^5$\,yr if the accretion rate is assumed to be
constant.

Fig.~\ref{infall} shows the CS(2--1) spectrum from Lee et~al.\
(\cite{lee99}) (telescope FWHM$\approx$27\arcsec\ at 86\,GHz), and our
DCO$^+$(2--1), H$^{13}$CO$^+$(1--0) and C$^{34}$S(2--1) spectra
(Juvela et~al.\ \cite{juvela02}) (FWHM$\approx$57\arcsec\ at
86\,GHz). Our spectra are at a single position nearest to the position
of the spectrum from Lee et~al.  The lines are not directly comparable
because the observations have different angular resolutions. The
CS(2--1) line is optically thick, while others are optically thin. The
optically thin lines peak at the absorption dip of the CS(2--1) line
at a velocity of about 2.5\,km\,s$^{-1}$, which confirms that the
double-peaked profile of CS(2--1) line is not due to two cloud
components at different velocities. In addition to the lines shown
here, the optically thick HCO$^+$(3--2) line shows a double-peaked
profile with a brighter blue peak, and the optically thin
H$^{13}$CO$^+$(3--2) line (Gregersen \& Evans \cite{gregersen00}) is
peaking at the absorption dip at the same velocity as the other
optically thin lines. This is consistent with the infall picture as
derived from the CS(2--1) observations.

The redshifted wing of the CS(2--1) line profile shown in
Fig.~\ref{infall} may be an indication of outflow.  Our
$^{13}$CO(1--0) observations (unpublished data) show many line
profiles with either blue- or red-shifted wings or shoulders. However,
the existence of blue- and red-shifted emission seems to be random,
which could be due to two overlapping outflows, both from FIR~1 and 2.

Juvela et~al.\ (\cite{juvela02}) have studied the virial equilibrium
of the L~183 cloud over a radius of $\sim 10$\arcmin\ ($\sim 0.3$\,pc)
by considering the kinetic and gravitational energies. They found that
the cloud as a whole is approximately in virial equilibrium. If,
however, the cloud is collapsing as suggested by asymmetric line
profiles, the equilibrium has been unstable to gravitational collapse.
This study of virial equilibrium can not be applied to the sources
IRS~1 and 2 because their radii are much smaller, $\sim 0.02$\,pc.

\subsection{Nature of the sources}

Based on a submillimetre continuum study of 21 starless cores,
including L~183, Ward-Thompson et~al.\ (\cite{wthompson94}) concluded
that the sources within cores are pre-protostellar rather than of
accreting Class~0 protostars. Firstly, the number of detected sources
was much greater than expected from the estimated lifetime of Class~0
objects. Secondly, the bolometric luminosities of the sources were so
low that if the luminosities were due solely to accretion, the ages of
protostars would be too small compared to the number of sources
detected.  This conclusion was made on statistical grounds and some of
the sources may indeed harbour a Class~0 protostar.

All the four sources in L~183 satisfy two defining properties of a
Class~0 protostar; 1) SED well fitted with a single modified blackbody
(although the temperatures are lower than those of Class~0 sources for
which T$_{\mathrm d} \approx 20-60$\,K (Andr\'e et~al.\
\cite{andreetal00})), 2) the ratio $L_{\mathrm{sub-mm}} /
L_{\mathrm{bol}} \gg 5 \times 10^{-3}$.  However, a genuine protostar
requires further indirect evidence for the presence of a central
protostellar object, such as a molecular outflow and/or a
cm-wavelength continuum source.  There is no confirmed outflow in
L~183 (see however Sect.~4.3).  All confirmed Class~0 objects, with
one exception, show manifestation of an outflow (Andr\'e et~al.\
\cite{andreetal00}). Harvey et~al.\ (\cite{harvey02}) have searched
for 3.6\,cm continuum emission from possible protostars in L~183 by
using the NRAO Very Large Array.  The half power diameter of the beam
was 5.3\arcmin\ (i.e.\ effective field of view) but the spatial
resolution of the VLA observations was $\sim 11$\arcsec. No source was
detected within the map of a size of about 8\arcmin square, centered
near FIR~1 and 2 (the sources FIR~3 and 4 are outside or at the very
edge of the map) with a 5-$\sigma$ limiting flux level of about
0.08\,mJy.  The seven Class~0 candidate sources detected in Bok
globules by Yun et~al.\ (\cite{yun96}) have 3.6\,cm continuum fluxes
between 0.2--5.3\,mJy.  It is thus most probable that FIR~1 and 2 are
pre-protostellar sources and not Class~0 objects.

The bolometric luminosities of FIR~1, 2 and 3 are very low,
$\sim$0.05\,L$_{\sun}$. In the list of confirmed Class~0 protostars by
Andr\'e et~al.\ (\cite{andreetal00}) the lowest L$_{\mathrm{bol}}$ is
0.15\,L$_{\sun}$.

\section{Summary and Conclusions}

The dark cloud L~183 has been mapped at far-infrared wavelengths with
the ISOPHOT instrument aboard ISO. Four unresolved sources were
detected in the cloud, called FIR~1, 2, 3 and 4, of which FIR~1 and 2
were previously known.  The SEDs of FIR~1 and 2 have been compiled by
combining our ISO fluxes at 120, 150 and 200\,$\mu$m with longer
wavelength fluxes from other studies. The SEDs are well fitted with a
single modified blackbody with dust temperatures T$_{\mathrm d}
\approx 8.3$\,K. The total masses (gas plus dust) of the sources are
$\sim$1.4\,M$_{\sun}$ and $\sim$2.4\,M$_{\sun}$, and the bolometric
luminosities are very low, $\sim$0.04\,L$_{\sun}$ and
$\sim$0.06\,L$_{\sun}$. Virial equilibrium consideration including
magnetic, kinetic, and potential energy and external pressure shows
that the sources have masses higher than or comparable to their virial
masses.

FIR~3 is detected at 120, 150 and 200\,$\mu$m.  Its dust temperature
is $\sim$13\,K and bolometric luminosity is
$\sim$0.03\,L$_{\sun}$. FIR~4 has been detected at 200\,$\mu$m only.

FIR~1 and 2 are probably gravitationally bound objects, rather
pre-protostellar sources than Class~0 protostars. The available data
do not provide definitive conclusions on the nature of the objects.

FIR~3 and 4 can be starless cold cores (either gravitationally bound
or unbound), pre-protostellar cores (gravitationally bound) or
protostars.  Virial equilibrium considerations cannot be made due to a
lack of accurate density structure and size information. Further
(sub)mm and centimetre continuum observations are required to study
the nature of the sources.

\begin{acknowledgements}
The work of K.L., K.M.\ and M.J.\ has been supported by the Academy of
Finland through grant Nos.\ 158300 and 173727.  We thank Carlos
Gabriel (ISO Data Centre, Villafranca) for providing us with the IDL
runtime version of PIA. We thank Derek Ward-Thompson for letting us
use the 450 and 850\,$\mu$m flux densities of FIR~2 before
publication.  ISOPHOT and the Data Centre at MPIA, Heidelberg, are
funded by the Deutsches Zentrum f\"ur Luft- und Raumfahrt DLR and the
Max-Planck-Gesellschaft. We thank the anonymous referee whose comments
substantially improved this article. 
\end{acknowledgements}

\end{document}